\documentclass[aps,prd,a4paper,onecolumn,amsmath,showpacs,superscriptaddress,nofootinbib,preprintnumbers,notitlepage,11pt]{revtex4-1}
\usepackage{amsmath,amssymb,latexsym,mathrsfs}
\usepackage{graphicx}
\usepackage{xcolor}
\usepackage{hyperref}
\usepackage{enumitem,booktabs,cfr-lm}
\usepackage[referable]{threeparttablex}
\usepackage{dcolumn}
\usepackage{bm}
\usepackage{multirow}
\usepackage{verbatim}

\renewlist{tablenotes}{enumerate}{1}
\makeatletter
\setlist[tablenotes]{label=\tnote{\alph*},ref=\alph*,itemsep=\z@,topsep=\z@skip,partopsep=\z@skip,parsep=\z@,itemindent=\z@,labelindent=\tabcolsep,labelsep=.2em,leftmargin=*,align=left,before={\footnotesize}}
\makeatother

\newcommand{\der}{\mathrm{d}}

\newcommand{\lcdm}{\Lambda\mathrm{CDM}}

\newcommand{\overbar}[1]{\mkern 1.5mu\overline{\mkern-1.5mu#1\mkern-1.5mu}\mkern 1.5mu}

\begin{document}

\title{Probing the Weak Gravity Conjecture in the Cosmic Microwave Background}

\author{Martin Wolfgang Winkler}
\affiliation{The Oskar Klein Centre for Cosmoparticle Physics, Department of Physics, Stockholm University, Alba Nova, 10691 Stockholm, Sweden}

\author{Martina Gerbino}
\affiliation{Istituto Nazionale di Fisica Nucleare (INFN), Sezione di Ferrara, Via Giuseppe Saragat 1, I-44122 Ferrara, Italy}
\affiliation{HEP Division, Argonne National Laboratory, Lemont, IL 60439, USA}

\author{Micol Benetti}
\affiliation{Dipartimento di Fisica  ``E. Pancini'', Universit\`a di Napoli  ``Federico II'', Via Cinthia, I-80126, Napoli, Italy}
\affiliation{Istituto Nazionale di Fisica Nucleare (INFN), Sezione di Napoli, Via Cinthia 9, I-80126 Napoli, Italy}

\begin{abstract}
The weak gravity conjecture imposes severe constraints on natural inflation. A trans-Planckian axion decay constant can only be realized if the potential exhibits an additional (subdominant) modulation with sub-Planckian periodicity. The resulting wiggles in the axion potential generate a characteristic modulation in the scalar power spectrum of inflation which is logarithmic in the angular scale. The compatibility of this modulation is tested against the most recent Cosmic Microwave Background (CMB) data by Planck and BICEP/Keck. Intriguingly, we find that the modulation completely resolves the tension of natural inflation with the CMB. A Bayesian model comparison reveals that natural inflation with modulations describes all existing data equally well as the cosmological standard model $\Lambda$CDM. In addition, the bound of a tensor-to-scalar ratio $r>0.002$ correlated with a striking small-scale suppression of the scalar power spectrum occurs. Future CMB experiments could directly probe the modulation through their improved sensitivity to smaller angular scales and possibly the measurement of spectral distortions. They could, thus, verify a key prediction of the weak gravity conjecture and provide dramatic new insights into the theory of quantum gravity.
\end{abstract}

\maketitle

\section{Introduction}
The cosmic microwave background (CMB) provides a window to physics at very high energy scales. It probes the era of cosmic inflation and could even contain imprints from the theory of quantum gravity. Within recent years, major theoretical advances have been made in understanding how such signatures could possibly look like. 
The progress roots in the observation that theories of quantum gravity have to fulfill certain self-consistency conditions. A particular intriguing example is the weak gravity conjecture (WGC) which -- in its original form -- constrains the strength of gauge forces relative to gravity~\cite{ArkaniHamed:2006dz}. The conjecture is motivated by the absence of an infinite tower of stable black hole remnants which would otherwise plague the theory.

Possibly even more important is the application of the WGC to non-perturbative axion physics~\cite{ArkaniHamed:2006dz,Rudelius:2014wla,delaFuente:2014aca,Rudelius:2015xta,Montero:2015ofa,Brown:2015iha}. This is because axions are the prime candidates to realize large-field inflation\footnote{While cosmological data do presently not require trans-Planckian excursions of the inflaton field, a natural choice of initial conditions and an observable tensor mode signal make large-field inflation a particular attractive candidate to describe the early expansion of the universe.} within a consistent theory of quantum gravity -- for which string theory is the leading candidate. String axions, descending from higher-dimensional p-form gauge fields, possess continuous shift symmetries which hold at all orders in perturbation theory and survive quantum gravity effects~\cite{Wen:1985jz,Dine:1986vd}. Non-perturbative instanton terms break the shift symmetries down to discrete remnants in a controlled way. In the simplest case, the resulting axion potential features the familiar cosine shape of natural inflation~\cite{Freese:1990rb}. Its periodicity is determined by the axion decay constant $f$ which -- in natural inflation -- has to be trans-Planckian since a too red spectrum of CMB perturbations would otherwise arise. While $f>1$ for a single fundamental axion does not arise in a controllable regime of string theory~\cite{Banks:2003sx,Svrcek:2006yi}, an effective trans-Planckian $f$ can consistently be realized via the alignment of two or more axions with sub-Planckian decay constants~\cite{Kim:2004rp}.

The WGC imposes that any axion must be subject to a (possibly subdominant) modulation with sub-Planckian periodicity $f_{\text{mod}}$. Natural inflation (with $f>1$) can still be realized through the axion alignment mechanism, but it necessarily comes with subdominant `wiggles' on top of the leading potential~\cite{Brown:2015iha,Brown:2015lia,Bachlechner:2015qja,Hebecker:2015rya,Kappl:2015esy}. The resulting scheme was dubbed `modulated natural inflation'~\cite{Kappl:2015esy}. It was realized that the modulations find a simple explanation in terms of higher instanton corrections. The non-perturbative breaking of the axionic shift symmetries in many cases comes from instantons which are described by modular functions (see e.g.~\cite{Dixon:1986qv,Hamidi:1986vh,Dixon:1990pc,Grimm:2007xm,Blumenhagen:2009qh,Grimm:2009ef,Arends:2014qca}). Therefore, the axion potential exhibits the desired cosine shape with additional wiggles which result from the higher harmonics in the modular functions. It is, in fact, not surprising that modular functions play a crucial role since -- as the WGC itself -- they are deeply connected to the duality symmetries of string theory. The explicit shape of the inflaton potential has been derived in~\cite{Kappl:2015esy},
\begin{equation}
  V=\Lambda^4\left(1-\cos\frac{\phi}{f}\right)\left(1-\delta_{\text{mod}}\,\cos\frac{\phi}{f_{\text{mod}}}\right)\,,
\end{equation}
where $\delta_{\text{mod}}$ is the amplitude of the modulation term with sub-Planckian periodicity $f_{\text{mod}}<1$. It is believed that this form of the potential applies to a wide class of large-field inflation models consistent with the WGC.

The purpose of the present work is to derive the signatures of the weak gravity conjecture in the CMB. The wiggles in the axion potential, which it requires, seed a characteristic modulation in the primordial power spectrum of scalar perturbations which is logarithmic in the angular scale. We will derive analytic expressions for the primordial power spectra and implement them in the Code for Anisotropies in the Microwave Background ({\sc CAMB})~\cite{Lewis:1999bs}. This will allow us to solve the Einstein-Boltzmann equations for cosmological perturbations and compute the CMB temperature and polarization power spectra. We will then directly test the modulation against the most recent CMB data by Planck~\cite{Aghanim:2019ame,Aghanim:2018eyx} and BICEP/Keck~\cite{Ade:2018gkx}. Finally, we will make exciting predictions for spectral distortions in the CMB which can be tested with future satellite missions.

\section{The Weak Gravity Conjecture}
In this section, we review the WGC~\cite{ArkaniHamed:2006dz} and its implications for large-field inflation in more detail. Originally, the WGC was formulated to constrain the strength of gauge forces relative to gravity. In its so-called electric version\footnote{There also exists a magnetic version of the WGC which states that any U(1) gauge theory with coupling $g$ breaks down at a cutoff scale $\Lambda < 1/g$~\cite{ArkaniHamed:2006dz}. This condition ensures that the  minimally charged magnetic object of the gauge theory is not a black hole. The application of the magnetic WGC to axion systems is less straight-forward compared to the electric version. In particular, it is still open, to which extent the magnetic WGC constrains effective trans-Planckian axion decay constants~\cite{Hebecker:2017wsu,Hebecker:2017uix,Reece:2018zvv,Henkenjohann:2019fdc}.} it states that any U(1) gauge theory coupled to gravity should contain a particle with charge-to-mass ratio\footnote{More precisely, $q$ in this expression stands for the product of charge and gauge coupling.} 
\begin{equation}\label{eq:wgc}
  \frac{q}{m}>1\,.
\end{equation}
In the absence of a lower bound on $q$, peculiar consequences would arise: for vanishing coupling strength, the gauge boson kinetic term becomes infinite. A non-propagating gauge boson implies that the gauge symmetry effectively behaves as a global symmetry. This appears problematic in the light of strong arguments against the existence of exact global symmetries~\cite{Banks:1988yz,Kallosh:1995hi,Banks:2010zn,Harlow:2018tng}. Violation of the weak gravity conjecture would, furthermore, imply that extremal black holes cannot decay. The theory would be plagued by an infinite tower of stable gravitational bound states. It was pointed out that such relics cause problems with the covariant entropy bound~\cite{Banks:2006mm}. Since there appear subtleties in the argument (see discussion in~\cite{Palti:2019pca}), the inconsistency of stable charged black holes is, however, not yet settled. 

There also exists a strong version of the weak gravity conjecture~\cite{ArkaniHamed:2006dz} (strong WGC) which insists that it is the lightest charged particle which must satisfy the condition $q/m > 1$. The stronger version is, however, most likely too restrictive. While it has originally been motivated in string theory, counter-examples have later been derived~\cite{Heidenreich:2015nta,Heidenreich:2016aqi}. It was also noted that even if a theory respects the strong WGC in the ultraviolet (UV), violations can occur in the effective low-energy theory obtained after Higgsing~\cite{Saraswat:2016eaz}.

In this work, the we are mainly interested in the application of the WGC to axion systems~\cite{ArkaniHamed:2006dz,Rudelius:2014wla,delaFuente:2014aca,Rudelius:2015xta,Montero:2015ofa,Brown:2015iha}. We consider an axion whose shift symmetry is broken to a discrete remnant via instanton terms. In the simplest case, the resulting axion potential takes the form
\begin{equation}\label{eq:natinflation}
  V=\Lambda^4\, e^{-\mathcal{S}} \left(1-\cos\frac{\phi}{f}\right)\,,
\end{equation}
where $\mathcal{S}$ denotes the instanton action, and we use Planck units ($M_P=1$) throughout this work. This is the familiar potential of natural inflation~\cite{Freese:1990rb}. Consistency with CMB constraints requires a trans-Planckian axion decay constant. While $f>1$ should not arise for a single fundamental axion~\cite{Banks:2003sx,Svrcek:2006yi}, an effective trans-Planckian $f$ can be realized by the interplay of two or more axions~\cite{Kim:2004rp}. It is exactly this possibility which is constrained by the WGC.

\begin{figure}[t]
\begin{center}   
 \includegraphics[width=14.5cm]{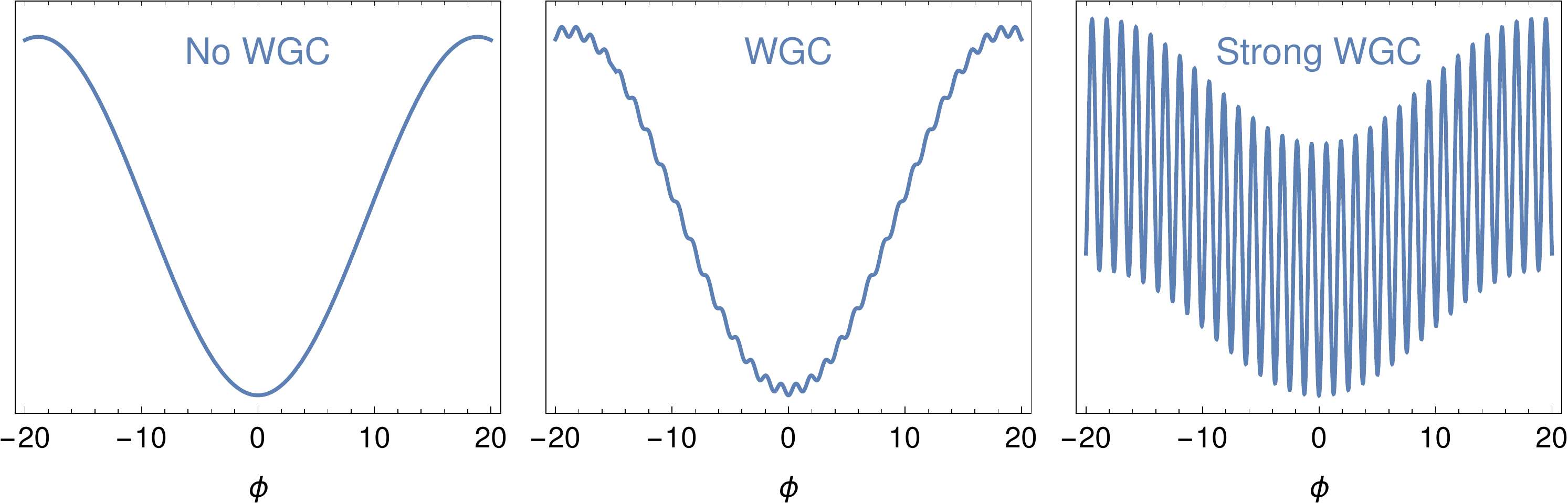}
\end{center}
\caption{Implications of the weak gravity conjecture for large-field inflation (schematic illustration of the potential). If the WGC is incorrect, standard natural inflation with a trans-Planckian axion decay constant can be achieved (left panel). If the WGC holds, large-field inflation can still be realized. However, a small-period modulation on top of the leading potential is required (middle panel). The resulting scheme is dubbed `modulated natural inflation'. The strong WGC (right panel) requires the small-period modulation to dominate such that inflation is completely spoiled.}
\label{fig:illustration}
\end{figure}

The string theory duality symmetries provide an inherent link between U(1) gauge theories and non-perturbative axion physics. In simple terms, the U(1) gauge charge translates to the inverse axion decay constant and the mass translates to the instanton action. For a single axion, the WGC then requires $(f\,S)^{-1}>1$ for at least one instanton of the theory. It is straight-forward to generalize the WGC to multi-axion systems with Lagrangian
\begin{equation}
\mathcal{L}= \frac{1}{2} \partial_\mu \phi_\alpha\partial^\mu\phi_\alpha - \Lambda_i^4\, e^{-\mathcal{S}_i}\Big[1- \cos{\left(c_{i,\alpha}\,\phi_\alpha\right) }\Big]\,,
\end{equation}
where the index $\alpha$ runs over the axions and $i$ runs over the instanton terms. The WGC is satisfied if the convex hull, spanned by the vectors $\pm (c_{i,\alpha}/\mathcal{S}_i)$, contains a ball with radius unity.\footnote{The convex hull condition has originally been formulated for the gauge theory version of the WGC in~\cite{Cheung:2014vva}.} More precisely, the required minimal radius is of $\mathcal{O}(1)$, but its exact value depends on the type of axion under consideration. Some examples are discussed in~\cite{Brown:2015iha}. In a system with many axions of different mass, the convex hull condition ensures that the one-axion WGC is satisfied after all heavier axions have been integrated out. The strong WGC imposes that the leading instanton(s) fulfill(s) the WGC.

One immediately realizes that the WGC has dramatic implications for large-field inflation which are illustrated in Fig.~\ref{fig:illustration}. The strong WGC entirely excludes natural inflation. This is because it requires a periodicity $f<1$ of the leading instanton term, i.e.\ of the one which dominates the potential. Given that the strong WGC is most likely overrestrictive, the more interesting scenario is, however, that only the `standard' WGC applies which is theoretically on much firmer grounds. In this case, the condition $(f\,S)^{-1}>1$ can be satisfied by a subleading instanton. Large-field inflation with $f>1$ can now be realized, but it necessarily comes with a subdominant high-frequency modulation on the potential. 

We expect that the amplitude of this modulation cannot be arbitrarily small. In the next section, we will discuss the explicit realization of trans-Planckian $f$ through the alignment mechanism. We will see that the modulation term increases rapidly for trans-Planckian $f$ and leads to an effective cutoff scale $\Lambda^4 < e^{-c\,f}$ with an $\mathcal{O}(1)$ number $c$. In the observationally most interesting regime $f=2-10$, the modulation can still be controlled, but it affects cosmological observables. This raises the exciting prospect of testing the WGC through CMB data. Before we discuss this in detail, we turn to the concrete realization of natural inflation with modulations.

\section{Modulated Natural Inflation}\label{sec:mni}

Natural inflation (NI) in its simplest form assigns a cosine potential to the inflaton (cf.~\eqref{eq:natinflation}). Let us briefly describe the axion alignment mechanism for realizing an effective trans-Planckian decay constant for the axion~\cite{Kim:2004rp}. One considers two canonically normalized axion fields $\phi_{1,2}$ with potential
\begin{equation}\label{eq:2axions}
V = \Lambda_a^4\, e^{-S_a}\left(1- \cos{\left[\frac{\phi_1}{f_1}+\frac{\phi_2}{f_2}\right] }\right)
+\Lambda_b^4\, e^{-S_b}\left(1- \cos{\left[\frac{\phi_1}{g_1}+\frac{\phi_2}{g_2}\right] }\right)\;,
\end{equation}
where all individual axion decay constants $f_{1,2},\,g_{1,2}$ are taken to be sub-Planckian. For simplicity, we consider the case $\Lambda_b^4\, e^{-S_b}\gg \Lambda_a^4\, e^{-S_a}$ such that we can integrate out the heavy axion direction $\tilde{\phi}\propto \tfrac{\phi_1}{g_1}+\tfrac{\phi_2}{g_2}$. The potential for the light axion $\phi\propto \tfrac{\phi_1}{g_2}-\tfrac{\phi_2}{g_1}$ is
\begin{equation}
V= \Lambda_a^4\, e^{-S_a}\left(1- \cos{\left[\frac{\phi}{f}\right] }\right)\,,\qquad f= \sqrt{g_1^2+g_2^2}\;\,\frac{f_1 f_2}{g_1 f_2-g_2 f_1}\,.
\end{equation}
The effective axion decay constant $f$ is strongly enhanced compared to the individual axion decay constants in the the alignment limit $f_1/f_2\simeq g_1/g_2$, allowing in particular for $f>1$.

However, we have seen in the previous section, that the pure cosine potential in combination with a trans-Planckian axion decay constant is in conflict with the WGC. Indeed, one easily verifies that the convex hull for the potential~\eqref{eq:2axions} becomes increasingly narrow in the alignment limit and does not contain the unit ball. Hence, the convex hull condition is violated. As described earlier, a subdominant modulation with sub-Planckian periodicity can reconcile NI with the WGC~\cite{Brown:2015iha,Brown:2015lia,Bachlechner:2015qja,Hebecker:2015rya}. From the bottom-up perspective, introducing the modulation may seem ad hoc. However, subleading instantons appear quite naturally in string theory. Non-perturbative terms often contain a series of higher harmonics. This is well-known for toroidal string compactifications~\cite{Dixon:1986qv,Hamidi:1986vh,Dixon:1990pc,Blumenhagen:2009qh}, but also established in more general string theory setups~\cite{Grimm:2007xm,Grimm:2009ef,Arends:2014qca}. The possible impact of higher instantons on inflation models has also been noted in~\cite{Abe:2014xja,Higaki:2015kta}.

The subleading instantons manifest in form of $\eta$- or $\theta$-functions which replace the exponential shape of a single instanton. As an example, we consider toroidal string compactifications with superpotential and Kahler potential~\cite{Kappl:2015pxa,Kappl:2015esy}
\begin{align}\label{eq:mnimodel}
  W &= \psi_1\: \Big(A_1\, \eta(T_1)^{2n_1}\eta(T_2)^{2n_2}-B_1\Big) +\psi_2\: \Big(A_2\, \eta(T_1)^{2m_1}\eta(T_2)^{2m_2}-B_2\Big)\;,\nonumber\\
  K &= |\psi_1|^2 + |\psi_2|^2 - \log(\bar{T}_1+T_1)- \log(\bar{T}_2+T_2)\,,
\end{align}
with the chiral superfields $\psi_{1,2}$ and the Kahler moduli $T_{1,2}$ which parameterize the volume of two internal sub-tori. The imaginary parts of the $T_i$ are identified with the two axions participating in the alignment mechanism. The parameters $A_{1,2}$, $B_{1,2}$, which are taken to be constants, descend from integrating out heavy chiral fields with non-vanishing vacuum expectation values. The coefficients $n_{1,2},\,m_{1,2}$ are determined by the localization properties of the chiral fields. The model above can e.g.\ be realized in heterotic orbifold compactifications, where the non-perturbative terms are identified with world-sheet instantons~\cite{Ruehle:2015afa}. By expanding the Dedekind $\eta$-function, one can verify that the superpotential contains an infinite number of subleading instantons,
\begin{equation}
  \eta(T)=e^{-\pi T/12} \times \prod\limits_{j=1}^\infty \left( 1 - e^{-2j\pi T}\right)\,.
\end{equation}
Including the full series of higher harmonics, it can be shown that the convex hull condition is satisfied even in the alignment case (see~\cite{Kappl:2015esy}).

The relevant potential of the light axion direction $\phi$ follows after setting the remaining fields to their vacuum expectation values and integrating out the heavy axion direction. We choose the indices $1,2$ such that $B_2>B_1$ and $T_{1,0}>T_{2,0}$, where $T_{i,0}$ denotes the expectation value of $T_i$. This leads to the approximate form of the potential~\cite{Kappl:2015esy}
\begin{equation}\label{eq:mnipotential}
  V=\Lambda^4\left(1-\cos\frac{\phi}{f}\right)\left(1-\delta_{\text{mod}}\,\cos\frac{\phi}{f_{\text{mod}}}\right)\,.
\end{equation}
Up to the last bracket, this is simply the potential of natural inflation with
\begin{equation}
\Lambda^4 \simeq \Lambda_a^4\, e^{-S_a} = \frac{A_1\,B_1}{2T_{1,0}T_{2,0}}
\,e^{-(n_1T_{1,0}+n_2T_{2,0})\,\pi /6}\,,
\end{equation}
and
\begin{equation}
  f=\frac{3\sqrt{2}}{\pi(n_1 m_2 - m_1 n_2)}\;\sqrt{\frac{m_1^2}{T_{2,0}^2}+\frac{m_2^2}{T_{1,0}^2}}\,.
\end{equation}
This part of the potential is obtained from the leading term in the expansion of $\eta(T_1)$ and $\eta(T_2)$. A trans-Planckian axion decay constant is realized under the alignment condition $n_1/ n_2 \simeq m_1/m_2$.

The last bracket in the expression~\eqref{eq:mnipotential} quantifies the deviation from natural inflation induced by the higher instantons in the $\eta$-series.\footnote{It is sufficient to include the next-to-leading terms in the $\eta$-expansion in order to arrive at the potential~\eqref{eq:mnipotential}. Contributions beyond this order are suppressed even in the alignment limit.} It arises in the form of a modulation on top of the leading potential with relative amplitude
\begin{equation}
\delta_{\text{mod}} \simeq 2 n_2 \,e^{-2\pi T_{2,0}}\,,
\end{equation}
and period
\begin{equation}
  f_\text{mod} = \frac{\sqrt{m_1^2+m_2^2}}{m_1}\; \frac{1}{2\sqrt{2}\pi T_{2,0}}\,.
\end{equation}
The relative modulation amplitude $\delta_{\text{mod}}$ increases towards the alignment limit. This is because the leading instantons partly cancel for alignment, whereas the higher orders are not affected by cancellation. Note that $f_{\text{mod}}$ is of the size of the original axion decay constants, i.e.\ it is generically sub-Planckian. The ratio $f/f_{\text{mod}}$ is a measure for the alignment -- the more the axion decay constants are aligned, the bigger this factor becomes.

A very interesting property of the model is that it exhibits an intrinsic relation between the alignment factor and the cutoff-scale. Among all possible parameter choices we find\footnote{We required $T_{1,2}>1$ and $A_{1,2},\,B_{1,2}<1$ in order to be in the controllable regime of the theory.}
\begin{equation}
  \Lambda^4 < \exp\left(-\frac{\pi}{36}\frac{f}{f_{\text{mod}}}\right),
\end{equation}
where the factor $\pi/36$ in the exponent is related to the expansion of the $\eta$-function and may somewhat differ for other modular functions. Since $\Lambda$ effectively determines the scale of inflation, it can be constrained observationally. The correct normalization of the CMB power spectrum requires $\Lambda>10^{-3}$ which implies
\begin{equation}\label{eq:maxalignment}
 \frac{f}{f_{\text{mod}}} \lesssim 10^3\,.
\end{equation}
The scheme described by the potential~\eqref{eq:mnipotential} will in the following be called `modulated natural inflation' (MNI). We point out that the shape of the potential does not depend on the concrete compactification, but merely on the fact that the higher harmonics exist. Presumably, it is applies universally to natural inflation models consistent with the weak gravity conjecture. 

In Tab.~\ref{tab:benchmark}, we provide an exemplary parameter choice for the MNI model. We will later show that this benchmark point provides a very good fit to existing CMB data.

\begin{table}[htp]
\begin{center}
  \begin{tabular}{|cccccccccc|cccc|ccccc|}
  \hline
  $n_1\!$ & $n_2$ & $m_1\!$ & $m_2$ & $A_1$ & $A_2$ & $10^4 B_1$ & $10^3 B_2$ & $T_{1,0}$ & $T_{2,0}$ & $10^2\Lambda$ & $f$ & $f_{\text{mod}}$ & $10^3\delta_{\text{mod}}$ & $10^9 A_0$ & $n_0$ & $\!-10^2 n_t$ & $\delta$ & $\Delta$\\ \hline
  $3$ & $7$ & $2$ & $5$ & $0.17$ & $0.30$ & $0.28$ & $0.70$ &  $1.96$ & $1.53$ & $0.34$ & $3.87$ & $0.198$ & $0.96$ & $1.862$ & $0.931$ & $0.119$ & $0.28$ & $-0.52$\\
  \hline
  \end{tabular}
\caption{Set of benchmark parameters for the model defined in~\eqref{eq:mnimodel}. Also shown are the derived potential parameters which enter~\eqref{eq:mnipotential} and the resulting power spectrum parameters (cf.~\eqref{eq:MNI_PR} and~\eqref{eq:MNI_PT}).}
\label{tab:benchmark}
\end{center}  
\end{table}

\section{Primordial Power Spectrum of Modulated Natural Inflation}

In this section we derive analytic expressions for the scalar and tensor power spectra of MNI. It is convenient to start with the simpler case of natural inflation, i.e. to set $\delta_{\text{mod}}=0$ for the moment. In a second step, we will later derive how the expressions are modified for non-vanishing $\delta_{\text{mod}}$. We introduce
\begin{equation}
    V_0= V\big|_{\delta_{\text{mod}}=0}=\Lambda^4\left(1-\cos\frac{\phi}{f}\right)\,.
\end{equation}
Furthermore, we define the (potential) slow roll parameters
\begin{equation}\label{eq:slowrollparameters}
  \varepsilon_0=\frac{1}{2}\left(\frac{\der V_0/\der \phi}{V_0}\right)^2\,,\qquad \eta_0=\frac{\der^2 V_{0}/\der\phi^2}{V_0}\,,
\end{equation}
where the index $0$ again indicates that we are neglecting the modulation for the moment. The scalar and tensor power spectra are well-approximated by a power law form determined by the slow roll expressions,
\begin{equation}
\mathcal{P}_{\mathcal{R},0}= A_0 \left(\frac{k}{k_*}\right)^{n_0-1}\,,\quad \mathcal{P}_{t,0}=-8 n_t A_0 \left(\frac{k}{k_*}\right)^{n_t}\,,
\end{equation}
where $k_*$ denotes the pivot scale. The normalization $A_0$ is given as
\begin{equation}\label{eq:A0}
A_0 \simeq \left.\frac{V_{0}}{24\pi^2\varepsilon_{0}}\right|_{\phi=\phi_*}\simeq \frac {\Lambda^4\left (\cosh\left[ \frac{N_*}{2 f^2} \right] + 
      4 f^2\sinh\left[ \frac{N_*}{2 f^2} \right] \right)^2} {24 \pi^2 f^2}\,.
\end{equation}
The scalar and tensor spectral indices are determined as
\begin{align}
  n_0&\simeq 1 - 6\varepsilon_0 + 2\eta_0 \Big|_{\phi=\phi_*}\simeq 1-\frac{\coth\left[\frac{N_*}{2 f^2}\right]}{f^2}\,,\label{eq:n0}\\
  n_t&\simeq -2\varepsilon_0 \Big|_{\phi=\phi_*} \simeq\frac{1-\coth\left[\frac{N_*}{2 f^2}\right]}{1+2 f^2}\,.\label{eq:nt}
\end{align}
In the above expressions, we have introduced $N_*$ which denotes the number of e-foldings when the CMB scales crossed the horizon. One typically finds $N_*\simeq 50-60$ with the exact value depending on the post-inflationary evolution of the universe. We have also employed that the corresponding field value $\phi_*$ is determined by
\begin{equation}\label{eq:efoldings}
  N_*=\int\limits_{\phi_{\text{end}}}^{\phi_{*}} \frac{\der\phi}{\sqrt{2\varepsilon_0}}\quad \Longrightarrow \quad \phi_*=2 f \arccos\left[\frac{e^{-N_*/(2 f^2)}}{\sqrt{1+\frac{1}{2 f^2}}} \right]\,,
\end{equation}
with $\phi_{\text{end}}$ marking the end of inflation (where the slow roll condition is violated).

We now turn to the power spectrum of MNI including the modulation. The slow roll parameters $\varepsilon$ and $\eta$ of the full model are introduced as in~\eqref{eq:slowrollparameters}, with $V_0$ replaced by $V$ (defined in~\eqref{eq:mnipotential}). Notice that each derivative acting on the modulated part of the potential pulls out a factor $f_{\text{mod}}^{-1} > 1 $. Therefore, the modulations have a much stronger impact on $\eta$ than on $\varepsilon$. In order to remain in the slow-roll regime, we need to require $|\eta|\ll 1$ which translates to
\begin{equation}\label{eq:slow}
   \delta_{\text{mod}} \ll f_{\text{mod}}^2\,.
\end{equation}
If this condition is violated, a strong scale-dependence of the power spectrum arises which is inconsistent with observation. The only caveat consists in very small $f_{\text{mod}}$, for which the power spectrum oscillations are so rapid that they are not individually traced in the CMB. This situation can, for example, arise in axion monodromy inflation models~\cite{Silverstein:2008sg,Flauger:2009ab}. However, in MNI such high-frequency oscillations are inaccessible due to the constraint on the alignment factor~\eqref{eq:maxalignment} which prevents too small $f_{\text{mod}}$. We expect not more than a few oscillations over the range of scales observable in the CMB. Therefore, we can safely require slow roll and impose~\eqref{eq:slow}. 

The condition~\eqref{eq:slow} ensures that we can treat the modulation as a perturbation in $\varepsilon$. Including terms up to first order in $\delta_{\text{mod}}$, we find
\begin{equation}
  \varepsilon \simeq \varepsilon_0 \left(1+\delta\sin\left[\frac{\phi}{f_{\text{mod}}}\right] \right)\,,
\end{equation}
where we introduced
\begin{equation}\label{eq:delta}
  \delta = \frac{\sqrt{2}\,\delta_{\text{mod}}}{f_{\text{mod}}\,\sqrt{\varepsilon_{0*}}}\,.
\end{equation}
Notice that we have replaced $\varepsilon$ by $\varepsilon_{0\,*}$ (= $\varepsilon_{0}$ evaluated at the pivot scale) in the definition of $\delta$. This is justified since it merely amounts to neglecting a second order correction (we fully keep track of the scale dependence of $\varepsilon_0$ in the leading term). The modulation in $\varepsilon$ causes a modulation in the scalar power spectrum
\begin{equation}\label{eq:fullscalarpower}
\mathcal{P}_{\mathcal{R}}\simeq \mathcal{P}_{\mathcal{R},0} \,\left(1-\delta\sin\left[\frac{\phi(k)}{f_{\text{mod}}}\right]\right)\,.
\end{equation}
As a final step, we need to relate the field value to a physical scale in the CMB. The modulation does (mildly) affect the relation $\phi(k)$. However, this is negligible compared to the direct impact of the modulation term in~\eqref{eq:fullscalarpower}. We will, therefore, employ the `unperturbed' relation~\eqref{eq:efoldings} and perform an expansion around $\phi_*$. Neglecting the scale-dependence of $\varepsilon$ in the vicinity of the pivot scale, we find $\phi=\phi_*+\sqrt{2\varepsilon_{0\,*}}\,(N-N_*)$. Taking into account that $k/k_*\simeq e^{-(N-N_*)}$, we finally arrive at
\begin{equation}
  \phi = \phi_* - \sqrt{2\varepsilon_{0\,*}}\,\log\frac{k}{k_*}\,,
\end{equation}
with $\phi_*$ determined from~\eqref{eq:efoldings}. Our final expression for the scalar and tensor power spectra of MNI reads
\begin{align}\label{eq:MNI_PR}
\mathcal{P}_{\mathcal{R}}&\simeq A_0 \left(\frac{k}{k_*}\right)^{n_0-1}\,\left(1-\delta\,\sin\left[\Delta-\frac{\sqrt{-n_t}}{f_{\text{mod}}}\,\log\frac{k}{k_*}\right]\right)\,,\\
\mathcal{P}_{t}&\simeq-8 n_t A_0 \left(\frac{k}{k_*}\right)^{n_t}\,,\label{eq:MNI_PT}
\end{align}
where we expressed $\varepsilon_{0\,*}$ in terms of the tensor spectral index $n_t$ and introduced
\begin{equation}\label{eq:Delta}
  \Delta =\frac{\phi_*}{f_{\text{mod}}}\,.
\end{equation}
Notice that we neglected modulations in the tensor power spectrum. The latter is set by the energy scale of inflation $V$, while the scalar power spectrum scales with $V/\varepsilon$. Therefore, the modulation in $\mathcal{P}_{t}$ is suppressed by $\delta_{\text{mod}}/\delta\sim (f_{\text{mod}} \varepsilon_{0\,*})^{-1}\ll 1$ compared to the one in $\mathcal{P}_{\mathcal{R}}$. Therefore, in the regime, where $\mathcal{P}_{\mathcal{R}}$ is consistent with observation, the modulations in $\mathcal{P}_{t}$ play no role.\footnote{The expressions~\eqref{eq:MNI_PR} and~\eqref{eq:MNI_PT} seem to suggest a (slight) violation of the inflationary consistency relation $r=-8n_t$ in MNI. In reality, the consistency relation is, however, satisfied due to a small modulation in $n_t$. The latter is neglected in~\eqref{eq:MNI_PT} since its effect on $P_t$ resides at the per mill level in the relevant parameter space of MNI.}

\begin{figure}[htp]
\begin{center}
\includegraphics[width=0.6\textwidth]{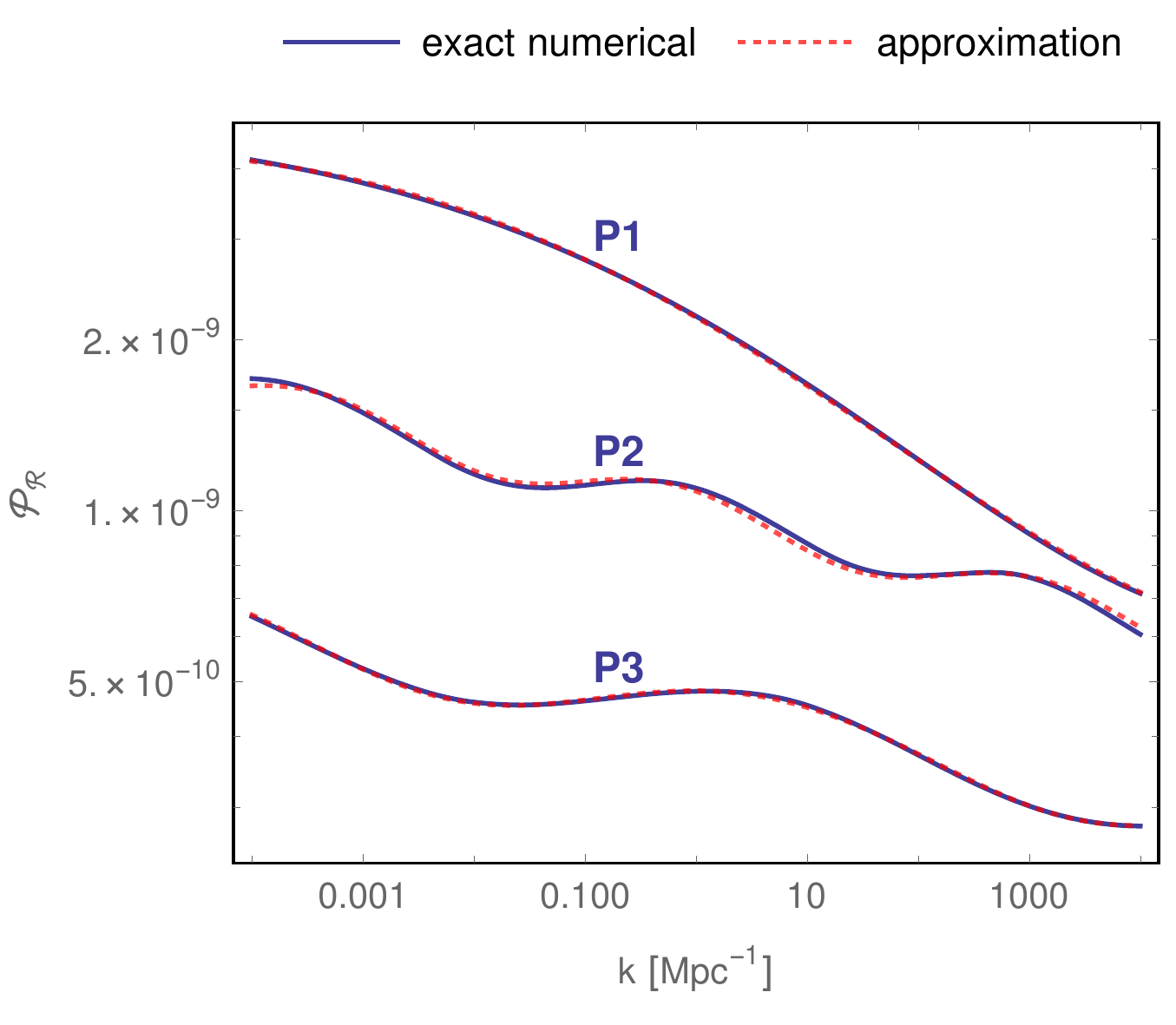}
\end{center}\caption{Primordial scalar power spectra for the parameter points listed in Tab.~\ref{tab:checkapprox}. Shown is the exact numerical solution based on the Mukhanov-Sasaki mode equations and the approximate analytic solution~\eqref{eq:MNI_PR}.}
\label{fig:checkapprox}
\end{figure}

In order to test the validity of our analytic approximation, we also performed exact numerical calculations of the primordial power spectra for a number of benchmark points. This was done by solving the full Mukhanov-Sasaki mode equations (as described in~\cite{Kappl:2015esy}). Figure~\ref{fig:checkapprox} compares the exact scalar power spectrum with our analytic approximation~\eqref{eq:MNI_PR} for the parameter points listed in Tab.~\ref{tab:checkapprox}. The spectra are in perfect agreement if we allow for a very mild refitting of the effective parameters $A_0$, $n_0$, $\delta$, $f_{\text{mod}}$, $\Delta$ compared to~\eqref{eq:A0},~\eqref{eq:n0},~\eqref{eq:delta} and~\eqref{eq:Delta}. The mismatch typically resides at the few per cent level\footnote{Only the phase $\Delta$ is not very accurately determined by our approximation~\eqref{eq:Delta}. This is expected since the phase is extremely sensitive to changes of the remaining parameters. We can safely ignore this issue since the phase is essentially a free parameter. Through very minor change of $f_{\text{mod}}$ one can realize any value of $\Delta$ while virtually not affecting the power spectrum otherwise.} as long as $\delta\lesssim 0.5$.

\begin{table}[htp]
\centering
\begin{tabular}{|c|ccccccccc|}
\hline
~P~ & $\Lambda$ & $f$ & $\delta_{\text{mod}}$ & $f_{\text{mod}}$ & $n_0$ & $-n_t\cdot 10^2$ & $A_0\cdot10^9$ & $\delta$ & $\Delta$\\
\hline
1 & $0.0033$ & $3.7$ & $0.0007$ & $0.16$ ($0.15$) & $0.925$ ($0.929$) & $0.09$ ($0.11$) & $2.25$ ($2.33$) & $0.29$ ($0.27$) & $-1.55$ ($-1.86$)\\
2 & $0.0042$ & $5.0$ & $0.0002$ & $0.08$ ($0.07$) & $0.952$ ($0.949$) & $0.39$ ($0.34$) & $1.18$ ($1.17$) & $0.08$ ($0.08$) & $1.42$ ($0.62$)\\
3 & $\,0.0043\,$ & $\;7.0\;$ & $\,0.0010\,$ & $\,0.20$ ($0.19$)$\,$~ & $\,0.963$ ($0.962$)$\,$ & $\,0.84$ ($0.83$)$\,$ & $\,0.50$ ($0.49$)$\,$ & $\,0.11$ ($0.13$)$\,$ & $\,0.96$ ($0.59$)~\\
\hline
\end{tabular}
\caption{Parameter sets leading to the primordial scalar power spectra of Fig.~\ref{fig:checkapprox}. For the parameters $A_0$, $n_0$, $\delta$, $f_{\text{mod}}$, $\Delta$ we first give the values derived from~\eqref{eq:A0},~\eqref{eq:n0},~\eqref{eq:delta},~\eqref{eq:Delta}. Refitted values obtained by matching the analytic approximation to the exact numerical power spectrum are given in brackets. The pivot scale was set to $k_*=0.05\:\text{Mpc}^{-1}$ and $N_*=60$.}
\label{tab:checkapprox}
\end{table} 

The deviation of $\mathcal{P}_{\mathcal{R}}$ from the standard power-law form results in an intriguing deviation from $\lcdm$ cosmology. If the modulations of the CMB power spectrum could be experimentally proven, this would yield strong indication that the WGC holds in nature. The prospect of testing general laws of quantum gravity by the spectra of cosmological perturbations is extremely exciting.

\section{CMB Analysis}\label{sec:method}
The analytic expressions of the primordial power spectra~\eqref{eq:MNI_PR} and~\eqref{eq:MNI_PT} can directly be implemented in standard tools for cosmological analysis and the corresponding predictions can be derived. We are particularly interested in the consistency of MNI with present CMB data and in the most striking observable differences compared to $\lcdm$. In the following, we describe the analysis method adopted in this work. The discussion of the results is given in Sec.~\ref{sec:results}.

\subsection{Monte Carlo analysis}\label{subsec:MCMC}
We perform a Monte Carlo Markov Chain (MCMC) analysis to explore the parameter space of MNI and to derive constraints on parameters from a combination of the latest cosmological data, employing the publicly available sampler {\sc CosmoMC}~\cite{Lewis:2002ah}. In order to solve the Einstein-Boltzmann equations for cosmological perturbations and compute the theoretical predictions
, such as the CMB temperature and polarization power spectra, we modified the current version of the Code for Anisotropies in the Microwave Background ({\sc CAMB})~\cite{Lewis:1999bs}, so that the primordial spectra\footnote{In the standard {\sc CAMB} code, the primordial power spectrum is parameterized, at first order, as the power-law $\mathcal{P}_\mathcal{R}=A_s {\frac{k}{k_{\ast}}}^{(n_s -1)}$, which we retain when we focus on constraints on the $\lcdm$ model. }
are now given as~\eqref{eq:MNI_PR} -~\eqref{eq:MNI_PT}. Since it is of particular interest, whether CMB data favor a non-vanishing modulation, we will treat the case $\delta=0$ separately. It corresponds to standard natural inflation and will, hence, be dubbed `NI' in the following. For comparison with the cosmological standard model, we will also perform the CMB analysis for the $\lcdm$ scenario (as reference model) with one parameter extension, i.e. leaving the tensor-to-scalar ratio as free parameter ($=\lcdm+r$).

The cosmological models are fully specified by the following set of parameters: the physical densities of cold dark matter $\Omega_c h^2$ and baryons $\Omega_b h^2$, the angular size of the sound horizon at recombination $\theta$, the reionization optical depth $\tau$, the primordial amplitude $\ln(10^{10} A_s)$ and the remaining inflationary parameters. For $\lcdm+r$, we vary the spectral index $n_s$ of scalar perturbations and the tensor-to-scalar ratio $r$. 

For the MNI model, the choice of input parameters is ambiguous: we could either express the model in terms of the potential parameters in~\eqref{eq:mnipotential}, or directly by the derived power spectrum parameters. In order to allow for a more meaningful statistical comparison with $\lcdm+r$, the input primordial power spectra should share a similar structure, i.e. we should treat $n_0$ on the same footing as $n_s$ in $\lcdm+r$. Therefore we decided in favor of the second option and to sample over $n_0$, $\delta$ and $f_{\text{mod}}$, $\Delta$ complemented with the constraint $n_0<1-\frac{2}{N_*}$ ($n_0<0.967$ for $N_*=60$).\footnote{The upper bound on $n_0$ is approached in the limit $f\rightarrow \infty$.} Notice that $n_t$ is not treated as an independent parameter. At each sampling step, we can invert~\eqref{eq:n0} to get $f$ and then obtain $n_t$ from~\eqref{eq:nt}. While~\eqref{eq:n0} cannot be inverted analytically, a very precise estimate is found by employing the Warner approximation for the inverse Langevin function. We obtain
\begin{equation}\label{eq:fapprox}
f\simeq\left(1-n_0-\frac{2}{N_*}\right)^{-1/4}\left(1-n_0+\frac{4}{N_*}\right)^{-1/4}\,.
\end{equation}
Due to the increased number of parameters and their mutual degeneracies, the MCMC analysis for MNI is extremely time-consuming. Therefore, we decided to fix the e-fold number $N_*=60$ which would correspond to the case of instantaneous reheating after inflation. We have run a test case to verify that our results are qualitatively unchanged if we allow $N*$ to float in the range $N_*=50-60$.\footnote{For the test case, the final distribution is somewhat peaked towards the boundary $N_*=60$, with negligible spread in the other parameters with respect to the case of fixed $N_*$.} In the special case of NI ($\delta=0$), the parameters $f_{\text{mod}}$ and $\Delta$ become irrelevant.

We impose the following priors to the MNI (and NI) inflationary parameters:
\begin{itemize}
\item linear prior on $n_0\in [0.8:0.967]$;
\item linear prior\footnote{We also tested a logarithmic sampling over $\delta$. The logarithmic prior tends to give more statistical weight to smaller values of $\delta$, therefore making the results less distinguishable from NI.} on $\delta \in [0: 0.5 ]$. The parameter $\delta$ describes a perturbation to the power-law scalar spectrum, so it must be $\delta<1$;
\item linear prior on $\Delta\in [-\pi: \pi ]$;
\item linear prior on $f_\mathrm{mod}\in [0.01: 0.5 ]$. The modulation frequency has to be sub-Planckian (see Sec.~\ref{sec:mni}).
\end{itemize}

In the analysis, we assume purely adiabatic initial conditions. For the concrete realization of modulated natural inflation described in Sec.~\ref{sec:mni} this amounts to neglecting quantum fluctuations in field directions orthogonal to the inflaton. This is a valid approximation in a large fraction of the model parameter space, where an effective single field description arises. We also include a contribution of $N_\mathrm{eff}=3.046$ active neutrinos, with a total mass of $\sum m_\nu=0.06\,\mathrm{eV}$.\footnote{In particular, we include one massive neutrino with mass $0.06\,\mathrm{eV}$ and $2.046$ massless species. This choice approximates the case of minimal mass in the normal ordering scenario for massive neutrinos, and has been adopted extensively by the Planck collaboration~\cite{Aghanim:2018eyx} and in the literature. A different choice for the parameterization of massive neutrinos and/or the presence of extra radiation in the early Universe might have an impact on the constraints on inflationary parameters, see e.g.~\cite{Gerbino:2016sgw,Archidiacono:2016lnv}. We defer the investigation of such effects to future works, as they are beyond the scope of this paper.}

We recall that sampling over the phenomenological parameters describing the primordial power spectra $\mathcal{P}_\mathcal{R}$ and $\mathcal{P}_t$ is only one of the possible choices for deriving constraints on inflationary parameters. For example, the Planck collaboration has extensively used the method of sampling the Hubble Flow Functions (HFF)~\cite{Stewart:1993bc,Gong:2001he,Leach:2002ar}. Here, we opt for the phenomenological approach of sampling over the power spectra parameters for two reasons: first, all the models under scrutiny in this work obey slow-roll conditions and therefore we do not need a method that allows for a more general description beyond slow-roll approximations. Secondly, it has been shown that the two methods (phenomenological and HFF) agree at a level that is acceptable to answer the question of whether the MNI model reasonably describes current cosmological data~\cite{Planck:2013jfk}. 

We choose to build our data set combining measurements of CMB temperature and polarization anisotropies as well as measurements of CMB lensing signal from the latest Planck satellite data release (the baseline combination labeled as ``TTTEEE+lowE+lensing'' in the Planck papers~\cite{Aghanim:2019ame,Aghanim:2018eyx}), measurements of the CMB degree-scale BB power spectrum from the BICEP/Keck collaboration (``BK15'')~\cite{Ade:2018gkx}, and measurements of the angular scale of the Baryon Acoustic Oscillations (BAO) from the SDSS-BOSS collaborations~\cite{Alam:2016hwk,Ross:2014qpa,6dF}. In addition to the aforementioned cosmological parameters, we also vary nuisance parameters describing foreground contamination to the cosmological signal, following the prescriptions adopted by the Planck and BICEP/Keck collaborations. 

We present the results of the analysis in Sec.~\ref{sec:results}. 

\subsection{Model comparison}\label{subsec:dic}
We also perform a model-comparison analysis of MNI with respect to the standard $\lcdm+r$ model and NI (i.e. MNI with fixed $\delta=0$). Indeed, in addition to providing constraints on the model parameters, we also want to know to what extent MNI describes the cosmological data as well as the standard $\lcdm+r$ scenario and whether a non-vanishing modulation is preferred (comparison with NI). To answer these questions, we employ the Deviance Information Criterion (DIC) as a statistical tool~\cite{DIC}. Given a certain model $\mathcal{M}$, the corresponding DIC is defined as
\begin{equation}\label{eq:dic}
DIC_\mathcal{M}\equiv-2\overbar{\ln\mathcal{L}(\theta)}+p_D\,,
\end{equation}
where the first term is the posterior mean of $\mathcal{L}(\theta)$, i.e. the likelihood of the data given the model parameters $\theta$, and the second term is the Bayesian complexity $p_D=2\overbar{\ln\mathcal{L}(\theta)}-2\ln\mathcal{L}(\tilde{\theta})$. The latter is a measure of the effective number of degrees of freedom in the model quantified as the difference in information content when the model is fitted with (pseudo)true parameters (represented by the posterior mean likelihood) and with an estimator of the true parameters ($\tilde{\theta}$). In what follows, we choose the best fit point $\tilde{\theta}\equiv\hat{\theta}$ as an estimator of the true parameters, although different choices (posterior mean, mode, medians) are equally valid. With the choice $\tilde{\theta}\equiv\hat{\theta}$, the model DIC can be rewritten as $DIC_\mathcal{M}=2\ln\mathcal{L}(\hat{\theta})-4\overbar{\ln\mathcal{L}(\theta)}$. The mean likelihood can be easily obtained from the output chains of the MCMC analysis. The best fit likelihood is computed separately for each model, employing the BOBYQA algorithm implemented in CosmoMC for likelihood maximisation.  

The choice of the DIC as a model-comparison criterion with respect to other statistical tools such as the Bayesian evidence and/or AIC (Akaike Information criterion)/BIC (Bayes Information Criterion) is dictated by the fact that it is not trivial to identify the correct number of effective degrees of freedom. Therefore, the implementation of AIC/BIC is not easily obtained. In addition, we have checked that the values of the Bayesian evidence of the models under scrutiny are very close to each other~\footnote{We obtained this result with MultiNest runs and Planck 2015 data. Given the high computational cost of implementing MultiNest analysis with complex cosmological models, and given that we do not expect the result to change qualitatively with Planck 2018 data, we have decided not to run MultiNest analyses with the latest Planck 2018 data.}. From the Bayesian evidence alone, we are not able to tell whether 
or not we are overfitting the data. Therefore, we need a measure of the effective number of parameters that a model can constrain. This measure is given by the Bayesian complexity $p_D$~\cite{Kunz:2006mc,Trotta:2008qt}. The DIC finally assesses the average performance of a model (given by the mean likelihood) with a penalty given by the Bayesian complexity. Equivalently, DIC measures the relative balance between the goodness of fit of a model (represented by the best fit likelihood) and the average performance of a model (represented by the mean likelihood). 

Models with lower DIC have to be preferred with respect to the reference value. It is always arbitrary to define the threshold for a model to be significantly preferred over another. In the case of DIC, an issue to take into account is the statistical noise introduced by the MCMC analysis and/or the minimization algorithm for the best fit. In other words, the $\Delta DIC=DIC_\mathcal{M}-DIC_{\mathcal{M}_{ref}}$ should be high enough that any statistical fluke can be considered negligible when assessing the model selection criterion. With this caveat, we follow previous works in literature and consider $\Delta DIC=10/5/1$ to provide, respectively, strong/moderate/null preference for the reference model.

\section{Results of the CMB Analysis}\label{sec:results}

The main results of our analysis are presented in Tab.~\ref{tab:Analysis}, where the constraints on the inflationary parameters are presented. Noteworthy, the ones on other cosmological parameter do not present significant differences with those of the standard model, so it was decided not to show them. To facilitate comparison, we also show the derived inflationary parameters, calculated at the pivot scale $k_*=0.05\,\mathrm{Mpc}^{-1}$. We employ the standard definitions $A_s=\mathcal{P}_{\mathcal{R}}(k_*)$ and
\begin{equation}
    n_s=1+\left.\frac{\der\log\mathcal{P}_{\mathcal{R}}}{\der \log k}\right|_{k=k_*}\,,\quad
    n_\mathrm{run}=\left.\frac{\der^2\log\mathcal{P}_{\mathcal{R}}}{\der (\log k)^2}\right|_{k=k_*}\,,\quad
    n_\mathrm{runrun}=\left.\frac{\der^3\log\mathcal{P}_{\mathcal{R}}}{\der (\log k)^3}\right|_{k=k_*}\,.
\end{equation}


\begin{table}[!]
\centering
\begin{tabular}[b]{|c|c|c|c|c|}
\hline
\multicolumn{1}{|c|}{ }&
\multicolumn{2}{c|}{ $\lcdm$ }& 
\multicolumn{2}{c|}{ NI }\\   
{Parameter}&
{\textbf{ mean}}& {\textbf{ best fit}}&
{\textbf{ mean}}& {\textbf{ best fit}}\\
\hline
$10^{9} A_s$
& $2.109\pm 0.030$ & $2.108$
& $2.096\pm 0.028$ & $2.093$
\\
$n_{s}$
& $ 0.9669\pm 0.0038 $ & $0.9671$
& $0.9612^{+0.0029}_{-0.0020}$ & $0.9621$
\\
$r$
& $ < 0.06$ & $0.02$
& $0.061^{+0.011}_{-0.015}$ & $ 0.064 $
\\
\hline
$n_t$
& $-0.0034^{+0.0030}_{-0.0014}$ & $-0.0025$
& $-0.0077^{+0.0019}_{-0.0014}$ & $ -0.0080$
\\
\hline
\hline
$\chi^2$
&$3547\pm8$& $3517$
& $3552\pm8$ & $3523$
\\
\hline
\end{tabular}
\quad
\begin{tabular}[b]{|c|c|c|}
\hline
\multicolumn{1}{|c|}{ }&
\multicolumn{2}{c|}{ MNI }\\    
{Parameter}&
{\textbf{ mean}}& {\textbf{ best fit}}\\
\hline
$10^{9} A_0$
& $1.93^{+0.18}_{-0.14}$ & $2.077$
\\
$n_{0}$
& $0.9498^{+0.0089}_{-0.0061}$ & $0.9514$
\\
$\delta$
& $[0.05,0.23]$ & $ 0.08$
\\
$\Delta$
& $-0.71^{+0.97}_{-0.86}$ & $- 0.20$
\\
$f_\mathrm{mod}$
& $ > 0.26$ & $ 0.30$
\\
\hline
$n_s$
& $0.9646\pm 0.0039$ & $ 0.9672$
\\
$n_\mathrm{run}$
& $ -0.0020 \pm 0.0037$ & $ -0.001$
\\
$n_\mathrm{runrun}$
& $-0.00062^{+0.00073}_{-0.00022}$ & $ -0.0006$
\\
$r$
& $0.0273^{+0.0074}_{-0.016}$ & $ 0.029$
\\
$n_t$
& $-0.0037^{+0.0022}_{-0.0011}$ & $ -0.0036$
\\
$f$
& $4.94^{+0.44}_{-0.86}$ & $4.88$
\\
\hline
\hline
$\chi^2$
& $3547\pm8$ & $3517$
\\
\hline
\end{tabular}
\caption{
Constraints and best fit values of the inflationary parameters in $\lcdm$, natural inflation (NI), and modulated natural inflation (MNI) models, using the combination of Planck 2018, BICEP/Keck and SDSS/BOSS data, see text for details. Constraints are $68\%$ C.L.. When only upper bounds can be placed, those bounds are 95\% C.L.. The scalar spectral index $n_s$ and the tensor-to-scalar ratio $r$ in the left table, and parameters in rows 4-8 are of the right table are computed at the pivot scale $k_*=0.05\,\mathrm{Mpc}^{-1}$. We neglected the very tiny running in NI in our parameterization of the power spectrum. The tensor spectral index $n_t$ is a derived parameters in the three models, with $n_t=-r/8$ in $\lcdm$ and $n_t$ given by~\eqref{eq:nt} in NI and MNI. In both tables, derived parameters are reported in the table sections below a horizontal line.}
\label{tab:Analysis}
\end{table} 

Turning first to natural inflation with the pure cosine potential, we confirm a mild tension with CMB data (in agreement with Planck results). In the parameter region, where NI reproduces the preferred spectral index, the tensor-to-scalar ratio is slightly to high. The best fit point features an axion decay constant of $f=6.8$ which minimizes the tension. It exhibits a smaller $n_s$, but larger $r$ compared to $\lcdm$.

\begin{figure}
\begin{center}
\includegraphics[width=0.5\textwidth]{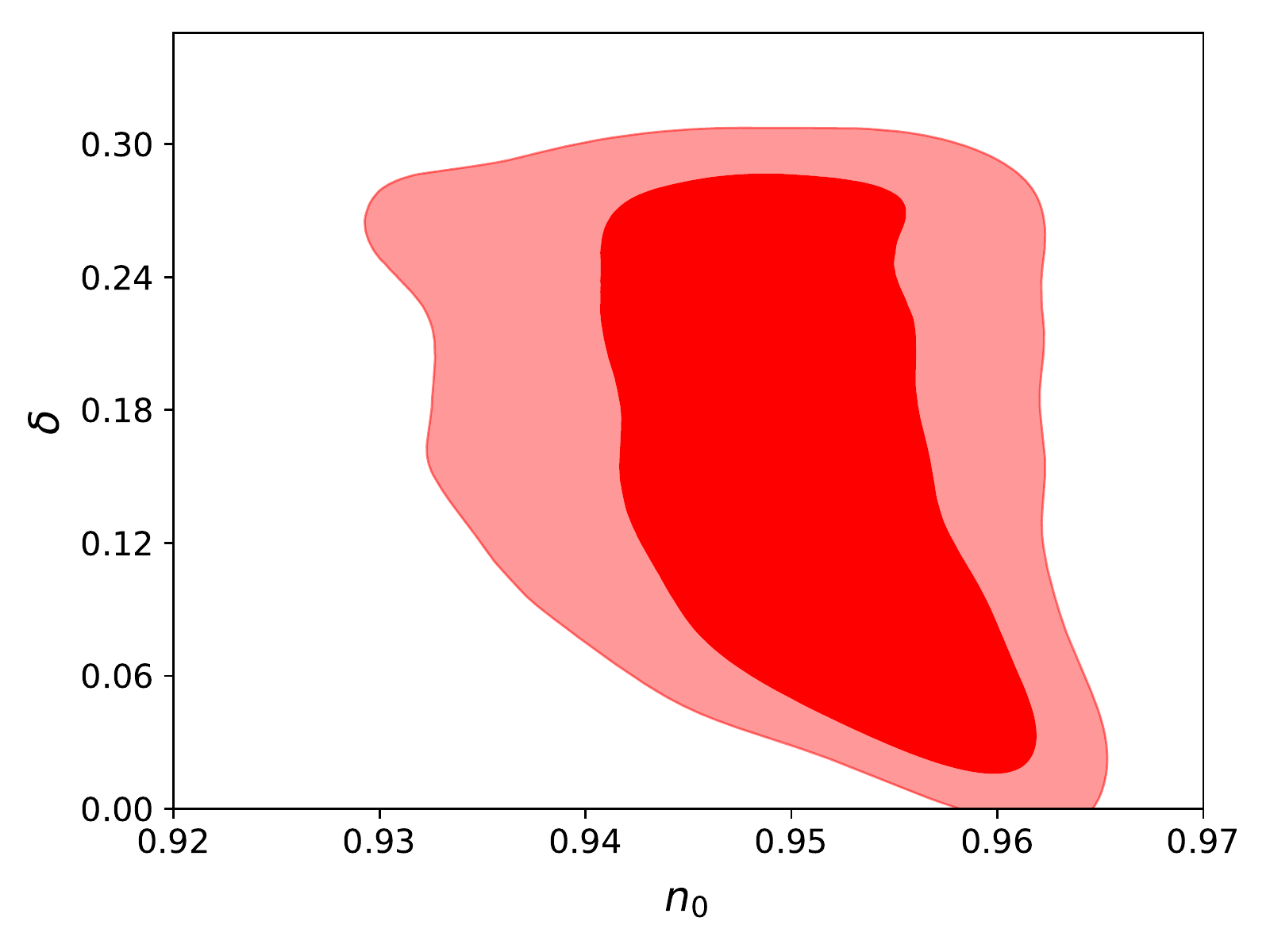}
\end{center}\caption{Two-dimensional contours in the $n_0$-$\delta$ plane for modulated natural inflation (MNI) model, for a combination of current cosmological data (Planck 2018, BICEP/Keck, BAO). Note the inverse degeneracy especially for smaller values of $\delta$: the MNI model can be a good fit to current cosmological data if the redder power-law term is compensated by the modulation term. However, current data already exclude values of $n_0$ that are too low.}
\label{fig:n0delta}
\end{figure}

We observe that MNI is able to resolve the tension faced by NI. This possibility is enabled by the modulation term in~\eqref{eq:MNI_PR} whose presence is enforced by the weak gravity conjecture.\footnote{Other possibilities to resolve the tension of natural inflation with CMB data have e.g.\ been discussed in~\cite{Peloso:2015dsa,Achucarro:2015caa}.} As stated in Tab.~\ref{tab:Analysis}, the data prefer a non vanishing modulation amplitude of $\delta=0.05-0.23$ (at the $1\,\sigma$ level). The key feature in MNI is the possibility to allow for smaller value of $n_0$, while keeping $n_s$ in the preferred window of $0.96-0.97$. Indeed, there is an inverse degeneracy between $n_0$ and $\delta$ as shown in Fig.~\ref{fig:n0delta}: when larger values of $\delta$ can be sampled, smaller values of $n_0$ are allowed. The modulation term adjusts such that it `compensates' the otherwise too strong scale dependence of the scalar power spectrum which would derive from the small $n_0$ (remember that in NI $n_0$ simply corresponds to $n_s$). A satisfactory description of CMB data is, however, only obtained within a limited range of $n_0\gtrsim 0.93$. For smaller values of $n_0$, one can still obtain $n_s$ in the desired range through a relatively large modulation term. However, the resulting power spectrum would deviate too strongly from the power-law form leading to a degraded fit.

Another interesting observation is that CMB data pin down the axion decay constant of MNI in a relatively narrow window $f\simeq 4.1-5.4$ (at the $1\,\sigma$ level). This implies that no large parametric enhancement of the axion decay constant is required to match CMB observations. Indeed, the typical alignment factor favored by CMB data is $f/f_{\text{mod}}=\mathcal{O}(10)$. This is well within the regime $f/f_{\text{mod}}<1000$ suggested by theory arguments (see Sec.~\ref{sec:mni}). In order to substantiate that the CMB preferred parameter range can be accessed in concrete UV models, we refer to our benchmark example of Tab.~\ref{tab:benchmark}. The corresponding scalar power spectrum is virtually indistinguishable from the best fit power spectrum within the range of scales probed by the CMB temperature data (see Fig.~\ref{fig:ps}).

\begin{figure}
\begin{center}
\includegraphics[width=0.45\textwidth]{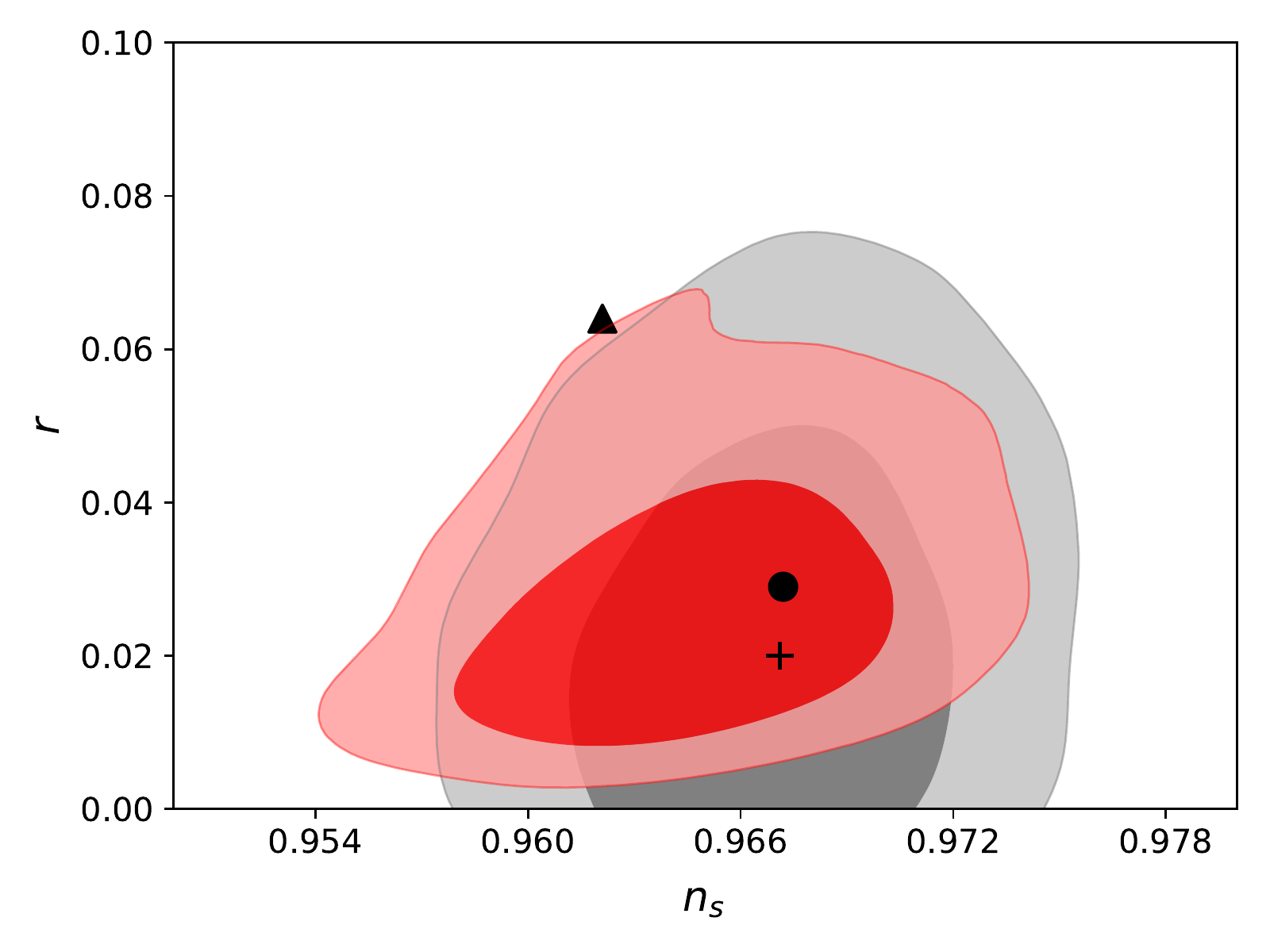}\includegraphics[width=0.45\textwidth]{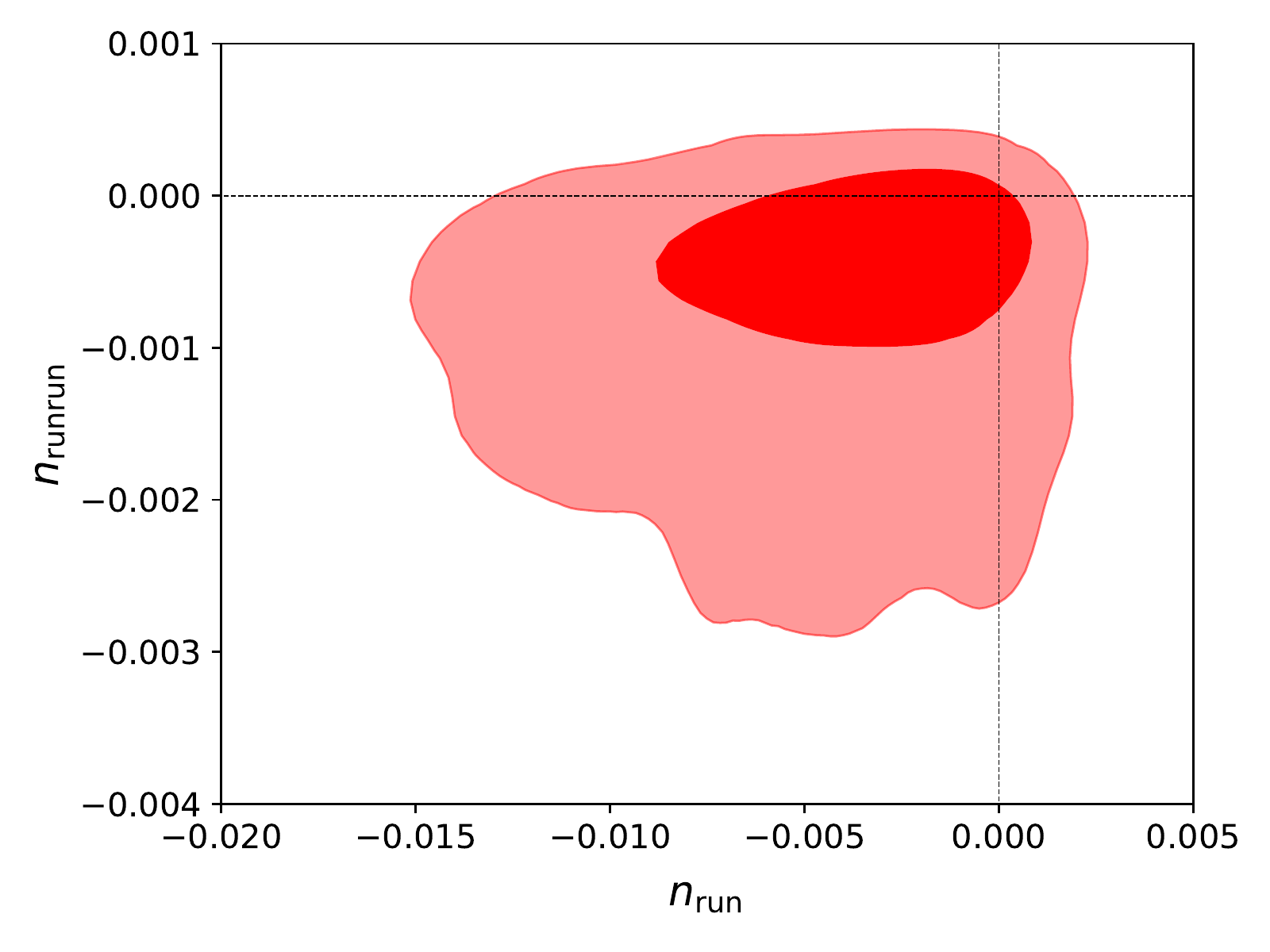}
\end{center}\caption{
\textit{Left}: Two-dimensional contours in the $n_s$-$r$ plane for a combination of current cosmological data (Planck 2018, BICEP/Keck, BAO). The red contours are for modulated natural inflation while the gray contours are for the $\Lambda$CDM+r model. The best fit points are shown as the black dot (MNI) and the black cross ($\lcdm+r$). For comparison we also show the best fit point for natural inflation (black triangle). The pivot scale is $k_*=0.05\,\mathrm{Mpc^{-1}}$. Note that the red contours close at 95\% C.L., i.e., vanishing $r$ is excluded at more than 95\% C.L. in the MNI model (see text for further details).
\textit{Right}: Two-dimensional contours in the
$n_\mathrm{run}$-$n_\mathrm{runrun}$ plane for modulated natural inflation, for the same combination of data as in the left panel. The horizontal and vertical dashed lines indicate the expected values of the two parameters in $\lcdm+r$, i.e. $n_{run}=0$, $n_{runrun}=0$. This point is at the border of the 68\% C.L. region in MNI.}
\label{fig:nsr-deltano}
\end{figure}

The two-dimensional probability contours in the $n_s$-$r$-plane are reported in the left panel of Fig.~\ref{fig:nsr-deltano}. The red contours are for MNI, while the gray contours refer to $\lcdm+r$. The preferred regions are similar for both models. We observe a very mild shift of the MNI preferred region towards smaller spectral index compared to $\lcdm+r$. However the $1\,\sigma$-regions strongly overlap. A more striking result concerns the tensor-to-scalar ratio. In the $\lcdm+r$ model, there is currently only an upper limit, $r<0.065$ at 95\% C.L.. The latter was obtained in the context of power-law primordial spectra with the inflation consistency relation $n_t=-8r$ imposed.

The situation changes drastically for MNI. As can be seen in Fig.~\ref{fig:nsr-deltano}, the probability contours in Fig.~\ref{fig:nsr-deltano} close at non-vanishing $r$. We derive the lower limit
\begin{equation}\label{eq:lowerlimitr}
  r> 0.002 \qquad\text{(at 99\% C.L. in MNI)}\,.
\end{equation}
If MNI is realized in nature, we hence expect a discovery of tensor modes by the next generation of CMB experiments. This is easy to understand: CMB constraints on the spectral index (and its scale dependence) exclude too small values of $n_0$ (see Fig.~\ref{fig:n0delta}). The lower bound on $n_0$ translates to a lower bound on $r$ through~\eqref{eq:fapprox} and~\eqref{eq:nt}. We emphasize that this lower limit is not driven by a preference of $r>0$ in the existing CMB data. Rather, it is forced by the model (in combination with the experimental constraints on $n_0$). We can conclude that the observation of a non-vanishing tensor mode signal is crucial for probing the weak gravity conjecture. If future experiments exclude $r> 0.002$, there would be little hope for observing any signature of the WGC in the CMB. If, on the other hand, a tensor mode signal is detected, experimental tests of the WGC may become feasible (see next section). These will be related to the scale-dependence of the spectral index which is visible in the right panel of Fig.~\ref{fig:nsr-deltano}, where we depict the probability contours for the running $n_\mathrm{run}$ and running-of-the-running $n_\mathrm{runrun}$ of the spectral index. Notice a preference for negative running in MNI.

We can now present the results of the model comparison analysis. We take $\lcdm+r$ to be the reference model, so that the $DIC_{\mathcal{M}_{ref}}\equiv DIC_{\lcdm}$ and $\Delta DIC_\mathcal{M}\equiv DIC_\mathcal{M}-DIC_{\lcdm}$. We obtain the following: 
\begin{itemize}
    \item $\Delta DIC_{\mathrm{MNI}}=0.5$, that is a no preference of the current cosmological data between $\lcdm+r$ and modulated natural inflation. Both models describe the existing data equally well;
    \item $\Delta DIC_{\mathrm{NI}}=3.8$, that is a mild preference of both, $\lcdm+r$ and modulated natural inflation with respect to natural inflation;
\end{itemize}

A deeper understanding can be obtained if we break down the $DIC$ values in their components, according to~\eqref{eq:dic}. The three models show the same Bayesian complexity $p_d\simeq30$, with differences at the level of the first decimal figure. Therefore, they share the same number of effective parameters needed to describe the data. The best fit and average likelihoods are also similar between $\lcdm+r$ and MNI, with $-2\ln\hat{\mathcal{L}}\simeq3517$ and $-2\overbar{\ln\mathcal{L}}\simeq3547$. Natural inflation is penalized with respect to both $\lcdm+r$ and MNI by significantly higher values of the best fit likelihood $-2\ln\hat{\mathcal{L}}=3523$ and of the average likelihood $-2\overbar{\ln\mathcal{L}}=3552$.

We should, hence, emphasize that the modulation term in the MNI potential fulfills two purposes: it renders the inflation model consistent with quantum gravity constraints (in the form of the weak gravity conjecture) and resolves the (mild) tension of NI with CMB data. 

\section{Future prospects}
In the previous section, we have seen that MNI provides as good as a fit to all current cosmological data as the cosmological standard model ($\Lambda\mathrm{CDM}+r$). In this section, we briefly discuss how future cosmological surveys can distinguish between the two models.

From the discussion in the previous section, we have seen that current data, when interpreted in the context of MNI, prefer non-zero values of $r>0.002$, see Fig.~\ref{fig:nsr-deltano}. Future CMB experiments will strongly improve the sensitivity to B-modes, and therefore will be beneficial to our understanding of the viability of the MNI model. For the reader's convenience, we briefly report the expected sensitivity $\sigma(r)$ on the tensor-to-scalar ratio $r$, or the expected exclusion limit at 95\% C.L.:
\begin{itemize}
    \item the ground-based Simons Observatory (first light in 2021)~\cite{Ade:2018sbj,Abitbol:2019nhf} will probe $\sigma(r)=0.003$.
    \item CMB-S4 (proposed project completion in 2029)~\cite{Abazajian:2019tiv,Abazajian:2019eic} will set a limit $r<0.001$ in absence of a tensor signal or clearly detect tensor modes if $r>0.003$.
    \item the satellite mission LiteBIRD (selected for launch in 2028)~\cite{Hazumi:2019lys} will reach $\sigma(r)\sim0.001$ (exact value depends on the specific noise model)
    \item the proposed satellite mission PICO~\cite{Hanany:2019wrm} would discover tensor modes at $5\sigma$ significance if $r>5\times 10^{-4}$. 
\end{itemize}
With future CMB surveys, we envisage that we could face two different situations: one possibility is that the overall amplitude of the primordial gravitational wave spectrum is so low that only an upper bound on $r$ will be put by future experiments. In this case, the probability contours for MNI in the $(n_s,r)$ plane will shrink, but they will be still centered at non-zero $r$ given the improved sensitivity of future cosmological surveys on $n_0$ (see the discussion in the previous section about the relation between $n_0$ and $r$ in MNI). Via the statistical tools employed in this work, one would find strong statistical preference for $\lcdm+r$ compared to MNI once the exclusions approach the lower bound set by data when interpreted in the context of MNI. At this point it would be clear that MNI is not the correct model of inflation and there is no hope of detecting signatures of the weak gravity conjecture in the CMB.

\begin{figure}
    \centering
    \includegraphics[width=.75\textwidth]{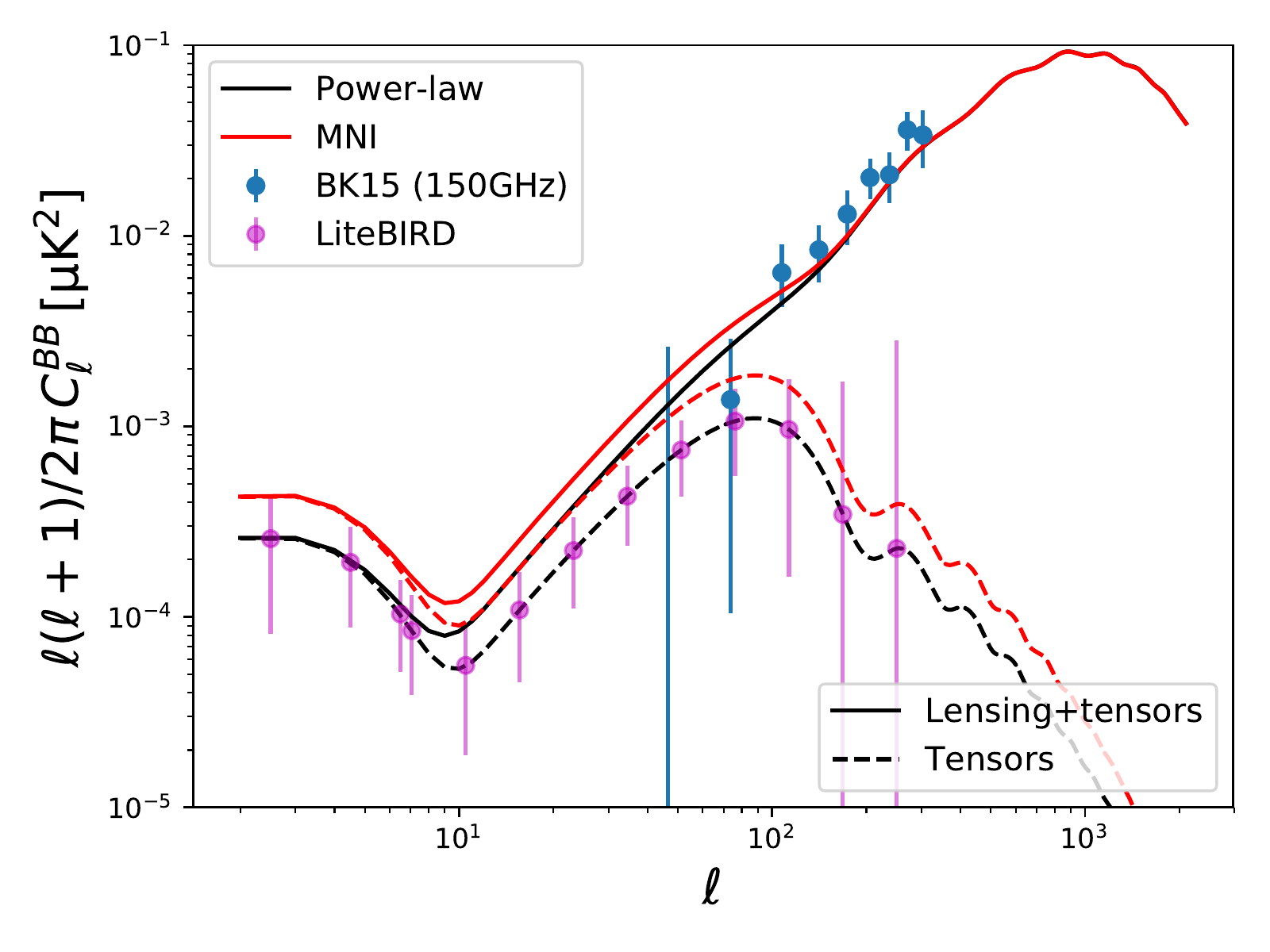}
    \caption{Best fit BB power spectra in $\lcdm+r$ (black lines) and modulated natural inflation (MNI) models (red lines), as found in this work for a combination of current cosmological data (Planck 2018, BICEP/Keck, BAO). The solid lines show the total signal, including lensing and tensor contributions. The dashed lines show the contribution from tensor modes only. The 68\% credible intervals for the CMB signal (lensing and possible tensor contribution) at the $150\,\mathrm{GHz}$ channel from the BICEP/Keck 2015 measurements are also reported as blue error bars, with the point marking the most probable value, or the 95\% upper limit with no point if the 68\% interval includes zero~\cite{Ade:2018gkx}. The pink error bars on top of the $\lcdm+r$ tensor spectrum are reported as an indication of the expected sensitivity from the future LiteBIRD satellite mission, and include cosmic variance, instrumental noise, and foreground residuals~\cite{Hazumi:2019lys,2019BAAS...51g.286L}.}
    \label{fig:BB}
\end{figure}

The much more exciting possibility is that future CMB experiments discover tensor modes at high statistical significance. Not only is the detection of non-vanishing $r$ a scientific milestone, it would also be an important step towards testing the MNI model and, thus, the possible signatures of the WGC. In Fig.~\ref{fig:BB}, we show the BB power spectra for both, $\lcdm+r$ and MNI, at the corresponding best fit points. The expected sensitivity from future surveys is represented with the error bars forecasted from LiteBIRD\footnote{We have chosen to show LiteBIRD as a reference case in Fig.~\ref{fig:BB} because it will have access to a wide range of angular scales from space with respect to ground-based experiments that are limited to the recombination bump ($\ell\sim80$). Nevertheless, we would like to remind the reader that the expected sensitivity from other future surveys would be able to detect a non-vanishing tensor signal roughly at the level of the lower bound on $r$ found in this analysis with current data.} as the pink errorbars on top of the tensor BB in $\lcdm+r$~\cite{Hazumi:2019lys,2019BAAS...51g.286L}. For comparison, the current error bars from BK15 are also shown~\cite{Ade:2018gkx}.\footnote{Data products available at \url{http://bicepkeck.org/bk15_2018_release.html}}
Notice that the expected BB signal is larger in MNI compared to $\lcdm +r$ due to the higher value of $r$ at the best fit point (see Tab.~\ref{tab:Analysis}). The magnitude of a detected tensor signal, in combination with sensitivity improvements on the spectral index, could thus already lead to a slight statistical preference for either MNI or $\lcdm +r$. We defer to future works for a thorough forecast analysis.

\begin{figure}[htp]
    \centering
    \includegraphics[width=.75\textwidth]{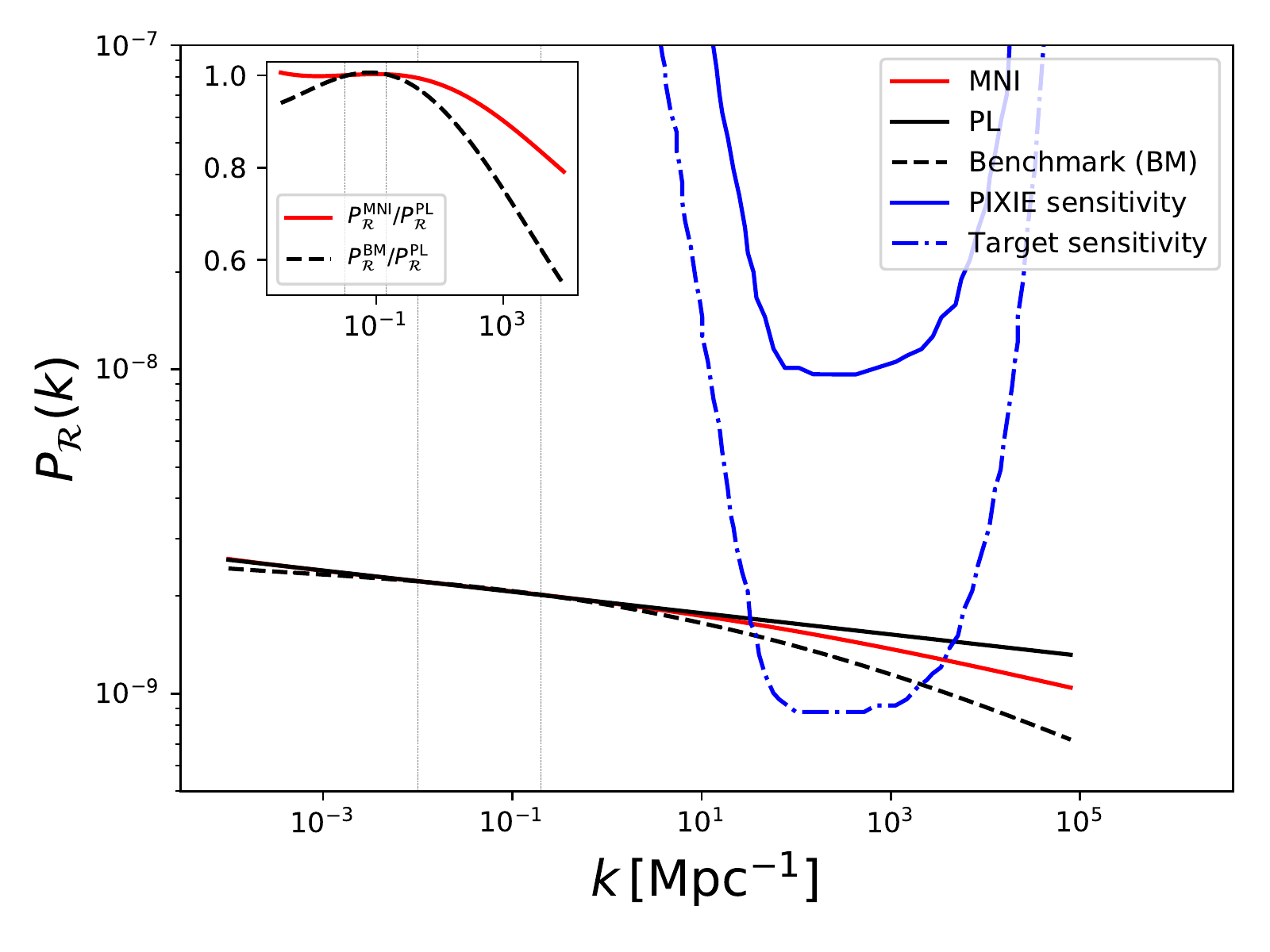}
    \caption{Primordial power spectrum of scalar perturbations $P_\mathcal{R}(k)$. We report the best fit power spectrum in $\lcdm+r$ (black line) and modulated natural inflation (MNI) model (red line) from a combination of current cosmological data (Planck 2018, BICEP/Keck, BAO). We also report the power spectrum as computed from the benchmark parameter choice listed in Tab.~\ref{tab:benchmark} (dashed black). The inset in the upper left corner shows the scalar spectrum ratio with respect to $\lcdm+r$ for MNI (red) and MNI-benchmark (dashed black). Current cosmological data tightly constrain the range of scales delimited by the two vertical lines, $0.01\,\mathrm{Mpc^{-1}}\lesssim k\lesssim 0.1\,\mathrm{Mpc^{-1}}$~\cite{Akrami:2018odb}. In this range, the three power spectra differ only at the sub-percent level. At small scales (large $k$), the MNI model predicts less power. Cosmological surveys targeting CMB spectral distortions could probe the range of scales where MNI currently predicts the largest deviations from the standard power-law spectrum, if the target sensitivity shown as the blue dot-dashed line can be reached. For comparison, we show the expected sensitivity from the proposed satellite mission PIXIE~\cite{Kogut:2011xw,10.1117/12.2231090}. Figure adapted from Refs.~\cite{Byrnes:2018txb,Chluba:2019kpb}.}
    \label{fig:ps}
\end{figure}

In addition to observable imprints on the tensor spectrum, a key feature of MNI is a scale-dependent modulation term in the spectrum of scalar perturbations. In Fig.~\ref{fig:ps}, we show the comparison between the best fit scalar power spectra obtained within $\lcdm+r$ (black line) and MNI (red line). We also show the power spectrum in the MNI benchmark model (see Tab.~\ref{tab:benchmark}) as the dashed black line. The vertical lines delimit the range of scales that can be accurately probed by current cosmological data, $0.01\,\mathrm{Mpc^{-1}}\lesssim k\lesssim 0.1\,\mathrm{Mpc^{-1}}$. Within this range, the MNI parameters can be arranged in such a way to recover an almost featureless power spectrum that deviates less than percent from the $\lcdm+r$ power spectrum. Within this range, a free-form Bayesian reconstruction of the power spectrum performed by the Planck collaboration~\cite{Akrami:2018odb} has shown that the (log) primordial spectrum can be recovered with a precision of a few percent. It is, however, clear that outside the range of scales probed by current CMB data, the uncertainty on a free-form reconstruction is much larger. Outside the CMB range, the difference between MNI and $\lcdm+r$ becomes relevant. In particular, the MNI spectrum features a considerable loss of power at small scales (large wave numbers $k$). This suggests that to probe with greater accuracy much smaller scales than those currently accessible can be the key to identify the modulation seeded by MNI.

The upcoming CMB experiments  Simons Observatory~\cite{Abitbol:2019nhf,Ade:2018sbj}, CMB-S4~\cite{Abazajian:2019tiv,Abazajian:2019eic} and possibly the proposed PICO~\cite{Hanany:2019wrm} satellite will increase the sensitivity to smaller angular scales (higher multipoles), and improve the constraints on the reconstruction of the scalar power spectrum with respect to the current sensitivity from Planck. If MNI is the correct model of inflation, these surveys could find indications of its scale-dependent spectral index. Large scale structure observations may offer another possibility to access the running of the spectral index favored by MNI (see e.g.~\cite{Stafford:2019flw}).

In addition to experiments aimed to measure CMB anisotropies, future surveys have been proposed that can measure spectral distortions in the CMB frequency spectrum (see e.g. the PIXIE proposal~\cite{Kogut:2011xw,10.1117/12.2231090}). These surveys would be able to probe much smaller scales ($k\simeq10^3\mathrm{Mpc}^{-1}$) than those accessible to experiments targeted to CMB anisotropies, and have the potential to test deviations from a standard power-law behaviour of $P_\mathcal{R}$~\cite{Byrnes:2018txb,Chluba:2019kpb}. It is clear that the possibility to access such a wide range of scales would either reduce the region of parameter space available to MNI to mimic a power-law behavior or allow to identify deviations from a standard power-law that manifest only at the smallest scales.

\begin{figure}
\begin{center}
\includegraphics[width=0.6\textwidth]{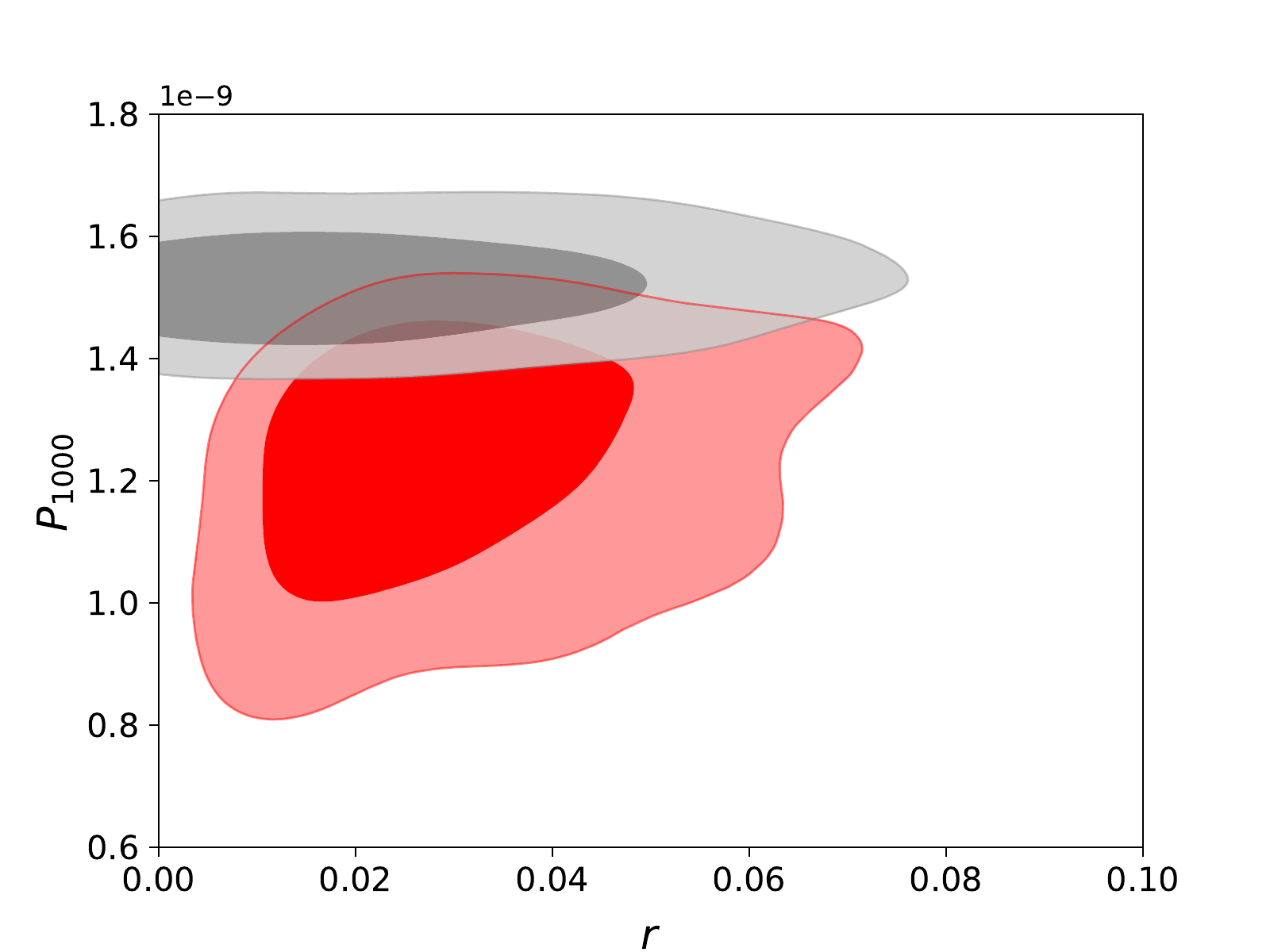}
\end{center}\caption{
Two-dimensional contours in the
$r$-$P_{1000}$ plane for modulated natural inflation (red) and $\lcdm+r$ (grey), where $P_{1000}$ is the scalar power spectrum at $k=10^3\,\mathrm{Mpc^{-1}}$. The contours are for a combination of current cosmological data (Planck 2018, BICEP/Keck, BAO). The pivot scale for the tensor-to-scalar ratio $r$ is $k_*=0.05\,\mathrm{Mpc^{-1}}$.}
\label{fig:p1000_r}
\end{figure}

In essence, we have identified a promising path to discover modulated natural inflation and, thus, to directly verify a key prediction of the weak gravity conjecture (namely the modulations in the power spectrum). Specifically, we have predicted a tensor mode signal correlated with a small scale suppression in the scalar power spectrum. In Fig.~\ref{fig:p1000_r}, we provide the posterior distribution in the two key observables: the tensor-to-scalar ratio and the amplitude of the scalar power spectrum at small ($k=10^3\:\text{MPc}^{-1}$) scales. A clear separation between the preferred regions is observed between MNI and $\lcdm+r$.

\section{Summary and Conclusions}
The weak gravity conjecture imposes stiking constraints on quantum field theories coupled to gravity. While the conjecture is supported by strong arguments related to black hole remnants, efforts towards its direct theoretical proof are still ongoing. In this work, we followed a different avenue and asked the question, whether predictions of the WGC can be tested experimentally. Specifically, we concentrated on signatures in the cosmic microwave background.

Our working assumption was that inflation is realized through an axion field and that its potential is generated by non-perturbative instanton effects. This framework is particularly appealing since the flatness of the inflaton potential carries a strong protection from the underlying shift symmetry which is preserved at the perturbative level. In the simplest case, the resulting potential would display a cosine shape as in the prominent model of natural inflation. The observed nearly scale invariant spectrum of inflationary perturbations requires the corresponding axion decay constant to be trans-Planckian.

The WGC provides concrete constraints on trans-Planckian axions. It states that a trans-Planckian axion decay constant can only be realized if the potential exhibits an additional (possibly subdominant) modulation with sub-Planckian periodicity which manifests in the form of `wiggles' on the leading potential. We provided a concrete realization of the modulation within string theory, where instanton effects arise in the form of modular functions. These contain a series of higher harmonics which can naturally be identified with the sub-Planckian modulation imposed by the WGC. The explicit inflaton potential, which is expected to hold rather generically in natural inflation models consistent with the WGC, is given in~\eqref{eq:mnipotential}. The model was dubbed `modulated natural inflation'.

We then derived analytic expressions for the scalar and tensor power spectra (see~\eqref{eq:MNI_PR} and~\eqref{eq:MNI_PT}). The wiggles in the inflaton potential imposed by the WGC translate to modulations in the scalar power spectrum. The latter exhibit a frequency which changes logarithmically on angular scales. The power spectrum is distinct from the cosmological standard model ($\lcdm+r$) in which it is restricted to a plain power law behavior. 

We implemented the primordial spectra of modulated natural inflation in the CAMB code and computed cosmological predictions such as the temperature and polarization power spectra. These were tested against the latest CMB data from Planck, BICEP/Keck and BAO data from SDSS-BOSS. Parameter constraints were derived in an MCMC analysis employing the sampler CosmoMC.

A first important result was that the modulations improve the consistency of natural inflation with the CMB. We then performed dedicated statistical comparison of modulated natural inflation against $\lcdm+r$. The outcome was that both models describe all existing cosmological data equally well. The Deviance Information Criterion yielded absolutely no statistical preference for any of the models. They also share very similar posterior distributions in the familiar $n_s$-$r$-plane (see Fig.~\ref{fig:nsr-deltano}).

A striking difference between both models is, however, that modulated natural inflation requires a non-vanishing tensor mode signal $r>0.002$ at 99\% C.L.. This lower limit is within reach of near-future ground-based CMB experiments such as the Simons Observatory. Furthermore, the scalar power spectrum of modulated natural inflation looks strikingly different from a power law outside the regime of scales presently probed by the CMB. 
In particular, a significant suppression of power at $k\gtrsim 100\:\text{Mpc}^{-1}$ is predicted (see Fig.~\ref{fig:ps}). Such small angular scales (large $k$) are accessible to proposed future CMB missions such as PIXIE via the measurement of spectral distortions. Although the expected signal is a factor of a few below the forecast sensitivity of PIXIE, we argue that the possibility to test the weak gravity conjecture via spectral distortions deserves further attention. Additional signatures related to the power suppression may arise e.g. in structure formation.

We can, hence, envision two situations: either CMB experiments will exclude $r > 0.002$ and rule out modulated natural inflation. This would blow all hopes of observing a signature of the WGC in the cosmic microwave background. If, on the other hand, a tensor mode signal is discovered, the weak gravity conjecture predicts that it must be intrinsically linked to a small scale suppression of the scalar power spectrum. 

The prospect of finding evidence for the weak gravity conjecture in future CMB data is extremely exciting. It would provide invaluable insights into the theory of quantum gravity. Provided that a tensor mode signal in the CMB is measured by near future CMB experiments, our analysis makes a poweful case for a dedicated satelite mission devoted to spectral distortion as a probe of signals generated by inflation.

\section*{Acknowledgments}
We would like to thank Katherine Freese, Massimiliano Lattanzi, Hans Peter Nilles and Luca Pagano for helpful discussions. Furthermore, we want to thank Marco Zatta for collaboration at an early stage of this project.
MWW acknowledges support by the Vetenskapsr\r{a}det (Swedish Research Council) through contract No. 638-2013-8993 and the Oskar Klein Centre for Cosmoparticle Physics.
MG acknowledges support by Argonne National Laboratory (ANL). ANL's work was supported by the DOE under contract DE-AC02-06CH11357.
MB acknowledge Istituto Nazionale di Fisica Nucleare (INFN), sezione di Napoli, iniziative specifiche QGSKY.
We also acknowledge the authors of the CosmoMC (A. Lewis) code. This work was developed thanks to the High Performance Computing Center at the Universidade Federal do Rio Grande do Norte (NPAD/UFRN) and the National Observatory (ON) computational support. 

\newpage

\bibliography{wgc}
\bibliographystyle{elsevier}
\end{document}